\documentclass[journal=jacsat,manuscript=article]{achemso}

\usepackage[version=3]{mhchem} 
\usepackage[T1]{fontenc}       

\usepackage[normalem]{ulem}
\usepackage{hyperref}
\usepackage{bm}
\usepackage{graphicx}
\usepackage{xfrac}
\usepackage{color}
\usepackage{amsmath}
\usepackage{amssymb}
\usepackage{epstopdf}
\usepackage{indentfirst}

\author{ Kaixuan Zhang$^{\dagger,\ddagger, \perp}$,Lin Li$^{\dagger,\ddagger, \perp}$,Hui Li$^{\dagger,\ddagger, \perp}$,Qiyuan Feng$^{\dagger,\ddagger}$,Nan Zhang$^{\dagger,\ddagger}$,Long Cheng$^{\dagger,\ddagger}$, Xiaodong Fan$^{\dagger,\ddagger}$, Yubin Hou $^{\dagger,\S}$,  Qingyou Lu$^{\dagger,\S,\parallel}$, Zhenyu Zhang$^{\ddagger}$,  and Changgan Zeng$^{\dagger,\ddagger,}$}
\affiliation{$^{\dagger}$Hefei National Laboratory for Physical Sciences at the Microscale (HFNL), CAS Key Laboratory of Strongly Coupled Quantum Matter Physics, and Department of Physics, University of Science and Technology of China, Hefei, 230026, China,
$^{\ddagger}$International Center for Quantum Design of Functional Materials, HFNL, and Synergetic Innovation Center of Quantum Information and Quantum Physics, University of Science and Technology of China, Hefei, 230026, China,
$^{\S}$High Magnetic Field Laboratory, Chinese Academy of Sciences, Hefei, 230031, China, and $^{\parallel}$Collaborative Innovation Center of Advanced Microstructure, Nanjing University, Nanjing, 210093, China}

\email{cgzeng@ustc.edu.cn}

\title[An \textsf{achemso} demo]
  {Quantum percolation and magnetic nano-droplet states in electronically phase-separated manganite nanowires}

\begin{document}

\begin{abstract}
One-dimensional (1D) confinement has been revealed to effectively tune the properties of materials in homogeneous states. The 1D physics can be further enriched by electronic inhomogeneity, which unfortunately remains largely unknown. Here we demonstrate the ultra-high sensitivity to magnetic fluctuations and the tunability of phase stability in the electronic transport properties of self-assembled electronically phase-separated manganite nanowires with extreme aspect ratio. The onset of magnetic nano-droplet state, a precursor to the ferromagnetic metallic state, is unambiguously revealed, which is attributed to the small lateral size of the nanowires that is comparable to the droplet size. Moreover, the quasi-1D anisotropy stabilizes thin insulating domains to form intrinsic tunneling junctions in the low temperature range, which is robust even under magnetic field up to 14 T, and thus essentially modifies the classic 1D percolation picture to stabilize a novel quantum percolation state. A new phase diagram is therefore established for the manganite system under quasi-1D confinement for the first time. Our findings offer new insight to understand and manipulate the colorful properties of the electronically phase-separated systems via dimensionality engineering.
\end{abstract}

\noindent KEYWORDS: Electronic phase separation, manganite nanowire, magnetic nano-droplets, tunneling, quantum percolation
\section{Introduction}

Shaping materials into reduced-dimensional structures offers additional flexibility to manipulate their physical properties, especially electronic transport properties. As for quasi-one-dimensional (quasi-1D) structures in uniform electronic states, high aspect ratio, large surface-to-volume ratio, or quantum confinement lead to a spectrum of intriguing effects, \emph{e}.\emph{g}., tunable bandgaps in semiconductor nanowires \cite{1}, and quantum interference in topological insulator nanoribbons \cite{2}. In particular, strong anisotropic magnetic and magneto-transport behaviors \cite{3,4,5,6}, as well as manipulation of individual skyrmions \cite{7} have been demonstrated in quasi-1D magnetic structures.

In contrast to the usually observed homogeneous electronic states, electronic phase separation (EPS) at mesoscopic scale can be stabilized in so-called electronically soft matters \cite{8}. A prototypical example is doped perovskite manganites, where antiferromagnetic charge-ordered insulating (COI) phase, ferromagnetic metallic (FMM) phase, and paramagnetic insulating (PI) phase may coexist in a certain range of temperature, due to the delicate coupling between the spin, charge, lattice, and orbital degrees of freedom \cite{9,10,11}. Dimensionality reduction has been adopted to further manipulate their physical properties. For example, spatial confinement engineering of phase-separated manganites, \emph{e}.\emph{g}., (La,Pr,Ca)MnO$_3$, effectively suppresses the number of transport paths, and develops interesting electronic transport phenomena, including giant resistance jumps \cite{12,13,14,15}, reentrant metal to insulator transitions \cite{16}, and intrinsic tunneling magnetoresistance \cite{17}, where the evolution of limited numbers of electronic domains plays critical roles.

Previous studies of manganites usually adopted top-down lithography technique to achieve spatial confinement \cite{12,13,14,15,16,17,18,19,20}. The very compound nature of manganites introduces complexities at the etched edges, which may affect their properties substantially \cite{19,20,21}. Here we prepare edge-free La$_{0.33}$Pr$_{0.34}$Ca$_{0.33}$MnO$_3$ (LPCMO)/MgO core-shell nanowires with superior structural quality by bottom-up method, which offer an ideal platform to investigate the intrinsic transport properties under quasi-1D confinement. We reveal that the quasi-1D confinement on the electronically phase-separated manganites is very sensitive to the order parameter fluctuations and even manipulates the phase stability, eventually leading to emergent physical effects.

\section{Results and discussion}

The single-crystalline LPCMO/MgO core-shell nanowires were grown by a two-step method following our previous work.\cite{22} The MgO nanowires were grown on MgO(001) substrates using chemical vapor deposition. The LPCMO shell layers were then deposited on the as-prepared MgO nanowires at $750\,^{\circ}\mathrm{C}$ using pulsed laser deposition. A repetition rate of 3 Hz was used, and the O$_2$ pressure was kept at 40 Pa during the deposition. After growth, the samples were annealed at the growth temperature with 1000 Pa O$_2$ pressure for 1 hour.

The morphology of the as-grown nanowires is shown in the scanning electron microscopy (SEM) image (Figure S1 in the Supporting Information). The length of the nanowires ranges between 1 and 10 $\mu$m. The core-shell structure, as well as the high single-crystalline quality with the absence of grain boundary, are clearly evidenced from the transmission electron microscopy (TEM) images in Figure 1a. The diameter of the MgO cores is about 20 nm, and the thickness of the LPCMO shells is around 20-30 nm. The slight thickness difference for the left and right sides of the LPCMO nanowire is due to that the two sides face oppositely during the growth.\cite{22,23} High-resolution TEM image in Figure 1a shows the LPCMO nanowires grow along the [001] direction, with a lattice constant of about 3.9 \AA, which is quite close to the bulk value of about 3.84 \AA. The LPCMO/MgO nanowires will be referred to as LPCMO nanowires hereafter for simplicity.

For LPCMO, FMM phase is revealed to dominate at low temperature \cite{9,24}. The aspect ratio of the LPCMO nanowires can be as high as 100, and thus should lead to giant magnetic shape anisotropy at low temperature. To avoid large paramagnetic signal from the MgO substrates, the nanowires were transferred onto a non-magnetic polydimethylsiloxane (PDMS) substrate. Figure 1b displays the typical zero field-cooled magnetic hysteresis loops for the LPCMO nanowires measured at 10 K. The apparent saturation moment is larger when the magnetic field is applied in-plane than out-of-plane. As discussed in the Supporting Information, such difference actually reflects the different degrees of magnetization when applying magnetic field along the two directions, suggesting magnetic anisotropy with easy axis along the nanowires. It is noted that the nanowires are oriented randomly on the substrate, so the magnetic anisotropy of a single nanowire should be far more significant than that observed for abundant nanowires (more details concerning the magnetic anisotropy are described in the Supporting Information).

The nanowires were then transferred onto SiO$_2$/Si substrates and patterned with the Cr/Au electrodes for transport measurements, with a typical device shown in Figure 1c. The magnetic anisotropy is also reflected by the magnetoresistance (MR, defined as [\emph{R}(\emph{H})-\emph{R}(\emph{H}=0)]/\emph{R}(\emph{H}=0)) as a function of the angle between the applied magnetic field and the nanowire long axis, as shown in Figure 1d. At 65 K, the MR value reaches maximum (-66.2\%) when the magnetic field is parallel to the nanowire and reaches minimum (-62.4\%) when vertical to the nanowire, which further validates the quasi-1D nature of the nanowires. Such quasi-1D confinement may further tune the physical properties substantially.

Next we focus on the temperature-dependent transport behaviors. Figures 2a and 2b show the resistance (\emph{R}) dependent on the temperature (\emph{T}) at different magnetic fields with the field parallel and perpendicular to the LPCMO nanowire, respectively. These \emph{R}-\emph{T} curves reveal exotic new features, which are absent for LPCMO bulk \cite{9}, thin films \cite{24} and strip structures \cite{12,13,14,15,16,17}.

First, a resistance kink is observed at \emph{T}* $\sim$ 200 K, and the rapid increasing trend of \emph{R} with decreasing \emph{T} is substantially suppressed below \emph{T}*, which strongly suggests the development of more conducting domains. Therefore the development of insulating states (for example, COI state and cluster glass state \cite{25}) can be excluded unambiguously. On the other hand, \emph{T}* is well above the Curie temperature (\emph{T}$_{\rm{c}}$ $\sim$ 150 K, which is determined from the \emph{T}-dependent magnetic moment (\emph{m}-\emph{T}) curve shown in Figure S4a), the resistance kink thus cannot be attributed to the onset of long-range ferromagnetic order.

For correlated electron systems, precursor phases with order parameter fluctuations may appear before the development of long-range-order phases. For example, the pseudogap phase could be the precursor of the global superconducting phase \cite{26}. As for the manganites, nanoscale droplet with short-range ferromagnetic interaction was proposed to emerge above \emph{T}$_{\rm{c}}$, which is the precursor of the FMM phase \cite{9,27,28}. Here the observed resistance kink at \emph{T}* therefore can be mainly attributed to the onset of such magnetic nano-droplet state. When \emph{T} decreases, the magnetic nano-droplets grow in size and eventually stabilize the FMM phase with long-range order below \emph{T}$_{\rm{c}}$. The application of magnetic field favors the growth of the magnetic nano-droplets and increases \emph{T}$_{\rm{c}}$ accordingly, as illustrated in Figure S4 and Figure 3a. On the other hand, \emph{T}* is almost independent on the magnetic field (Figure 2 and Figure 3a), suggesting that the onset of the magnetic nano-droplet state is insensitive to the applied magnetic field.

Such resistance kink is absent for the LPCMO bulk, thin films or strips, and could be ascribed to the fact that the size of the magnetic nano-droplets is much smaller than the lateral size of these samples, and their contribution to the electronic transport is therefore negligible. In sharp contrast, for the LPCMO nanowires with extreme aspect ratio, the size of the magnetic nano-droplets is comparable to the lateral size, and their appearance essentially modifies the transport behavior, manifested as a kink in the \emph{R}-\emph{T} curves.

Below \emph{T}$_{\rm{c}}$, FMM phase emerges, coexisting with the insulating phases, developing an EPS state, in which the FMM phase grows in proportion with further decreasing \emph{T}. For the LPCMO nanowires, the wire width (below 100 nm) is much smaller than the size of the electronic domains in the EPS state (submicrometer) \cite{9,29,30}, and each metallic or insulating domain thus spans the width of the nanowires to connect each other in series. As shown in Figure S5, the \emph{R}-\emph{T} curve below \emph{T}* can indeed be well fitted by a series two-resistor model (more details are described in the Supporting Information), validating that only a single transport path instead of transport network exists along the LPCMO nanowire.

It is noted that the fitting curve substantially deviates from the experimental one below a critical temperature \emph{T}$_{\rm{qp}}$ ($\sim$65 K). An apparent insulator-metal transition at \emph{T}$_{\rm{qp}}$ is featured for zero or low magnetic fields, while the residual resistance remains extreme high at low temperature. When the magnetic field is higher than 6 T, the apparent insulator-metal transition disappears, and a high-resistance plateau develops in the low temperature range below \emph{T}$_{\rm{qp}}$. These behaviors differ essentially from that of LPCMO thin films, where only a typical insulator-metal transition shows up with low residual resistance at low temperature (Figure S6a). It is also noted that \emph{T}$_{\rm{qp}}$ decreases with increasing magnetic field for the LPCMO nanowires, while the insulator-metal transition temperature increases with increasing magnetic field for the LPCMO thin films (Figure S6a), further suggesting that they have different origins. The observed high-resistance plateau is also different from the behavior of insulators, which shows exponentially increasing resistance with decreasing \emph{T}.

For LPCMO thin films, as \emph{T} decreases, the insulating phases is getting more and more energetically unfavorable, and the fraction of the FMM phase increases continuously at the expense of the insulating phases. When it is above the percolation threshold (about 25\% and 50\% for three-dimensional simple cubic lattice and two-dimensional square lattice, respectively \cite{31}), percolative conducting paths form through the connected FMM domains, manifested as an insulator-metal transition, leading to a low residual resistance \cite{24,29}.

For a LPCMO nanowire with extreme aspect ratio, however, only a single transport path exists as discussed earlier, and thus each domain should contribute to the transport substantially. In the classic 1D percolation picture, percolative conduction develops only when the insulating regions disappear completely, namely, the percolation threshold for the FMM phase is 100\% for such quasi-1D transport. Therefore at low temperature, the carriers still have to go through the much reduced insulating domains, although the FMM phase dominates the nanowires. Strikingly, when an insulating domain is shrunk to be thin enough, it may act as an intrinsic tunneling barrier to allow electron tunneling between its two neighboring FMM domains.

It has been shown that among the three Rowell criteria commonly adopted to identify tunneling junctions, only the weak insulating-like temperature dependence of the resistance remains a solid criterion \cite{32}. As seen in Figure 2, all the observed high resistance plateaus under magnetic fields higher than 6 T indeed exhibit weak insulating behavior. The \emph{R}-\emph{T} curves below \emph{T}$_{\rm{qp}}$ measured at 6 T, 9 T and 14 T were then fitted using the well-established tunneling equation: $R = \;\frac{W}{{1 + {{\left( {\frac{T}{{{T_0}}}} \right)}^2}}}$ \cite{33}, where \emph{W} and \emph{T}$_0$ are the fitting parameters. As plotted in Figure 2c, all these experimental curves can be well fitted in the low temperature range, further verifying that well-defined tunneling junctions are formed under high field below \emph{T}$_{\rm{qp}}$.

The \emph{T}-dependent transport behaviors under different magnetic fields can be understood as follows. For all cases, \emph{T}$_{\rm{qp}}$ signifies the onset of quantum tunneling, instead of an insulator-metal transition. For zero or low magnetic fields, only a very small fraction of the insulating domains are thin enough to act as tunneling barriers at \emph{T}$_{\rm{qp}}$ (see Figure 3b), which can be further verified from the weak tunneling MR effect obtained in the low temperature regime (see Figure S7 and detailed discussions in the Supporting Information). The remaining insulating regions further convert into the FMM phase with decreasing \emph{T}, leading to a decreasing resistance with decreasing \emph{T} below \emph{T}$_{\rm{qp}}$. While for magnetic fields higher than 6 T, the insulating phases are effectively suppressed by the high field, and only survive as a new type of domain walls between FMM domains to act as well-defined tunneling barriers at \emph{T}$_{\rm{qp}}$ (also see Figure 3b), resulting in the observed high-resistance plateaus below \emph{T}$_{\rm{qp}}$.

The survival of the insulating phases below \emph{T}$_{\rm{qp}}$ even under high magnetic fields is also confirmed from the magnetic measurements. Figures S4a-f show the \emph{m}-\emph{T} curves under different magnetic fields. The zero-field-cooling (ZFC) curve and the field-cooling (FC) curve deviate below a blocking temperature \emph{T}$_{\rm{b}}$. Such deviation identifies a blocked state with EPS \cite{34}. As summarized in Figure 4a, \emph{T}$_{\rm{b}}$ decays slowly in the high field region, and is about 20 K even under field up to 7 T. This observation suggests that the insulating phases are robust against applied magnetic field in the LPCMO nanowires, and can serve as tunneling barriers when they are shrunk to be thin enough. In contrast, \emph{T}$_{\rm{b}}$ almost goes to zero under high magnetic field for the LPCMO thin films (see Figure 4a and Figures S6b-g, which implies the complete disappearance of the insulating phases under high magnetic field in the thin films.

The distribution of the magnetic domains was further directly imaged using a home-made magnetic force microscopy (MFM) \cite{35,36}. Figure 4b shows the topography image of a LPCMO nanowire with length of about 1 $\mu$m, where the width is broadened by tip geometric effect. The corresponding MFM images under different applied magnetic fields scanned at 50 K are displayed in Figures 4c,d and more in Figures S8b-j. It is clear that in response to the increasing magnetic field, the FMM domains grow in size at the expense of the insulating phases. Nevertheless, the insulating phases still survive even under magnetic field as high as 3 T, as evidenced from the contrast variation in Figure 4d and the corresponding line profile shown in Figure 4g.

It is noted that there is no direct correlation between the MFM images and the topology image. Moreover, the locations of the insulating domains and their distances change with magnetic field. When the field is altered from 0.03 T to 3 T, for example, the distance variation between the two insulating domains indicated by the arrows shown in Figures 4f and 4g is about 100 nm. These observations strongly suggest that the robustness of the insulating phases under high magnetic field in the low temperature range is not originated from defect pinning, and collectively point to an intrinsic origin of quasi-1D confinement.

The apparent insulator-metal transition with high residual resistance at low temperature has also been observed for LPCMO submicrometer-wide bridges, and was also attributed to the tunneling across the insulating barriers \cite{17}. However the insulating barriers are metastable in the bridges, and disappear upon application of a relatively weak magnetic field (below 1 T) \cite{17}. In sharp contrast, the insulating barriers are so robust and survive even under magnetic field up to 14 T for the LPCMO nanowires.

It is noted that no giant resistance jump is observed in the \emph{R}-\emph{T} curves (Figure 2). Correspondingly, the growth of the FMM domains and the shrink of the insulating domains in size only evolve gradually from the magnetic-field-dependent MFM images. In contrast, giant resistance jump was generally observed for the micrometer-wide strips or submicrometer-wide bridges, which was attributed to a sudden phase conversion of a single domain from insulating to FMM \cite{13,14}. Our observations further support the robustness of the insulating phases in the LPCMO nanowires with extreme aspect ratio.

We also performed \emph{R}-\emph{T} measurements for several other LPCMO nanowires, and the results shown in Figure S9 reveal the same features as in Figure 2, suggesting that the properties discussed above are universal for the LPCMO nanowires.  On the other hand, strain between the LPCMO overlayer and the MgO template may also play a role to determine the observed properties. From the TEM characterization (Figure 1a), the lattice spacing of the LPCMO part is about 3.9 \AA, close to the bulk lattice spacing of about 3.84 \AA, while significantly different from the MgO lattice spacing of about 4.2 \AA. Moreover, Moir\'{e} pattern is also observed, further confirming the lattice difference between the LPCMO overlayer and the MgO template. These results suggest that the stress induced by the lattice mismatch is quickly released in the initial few LPCMO layers, and cannot account for the observed novel phenomena in the LPCMO nanowires.

Therefore the dimensionality should be mainly responsible for the observed effects, especially the robust quantum tunneling across the intrinsic insulating barriers. It has been proposed theoretically that in an EPS system such as manganite, two ferromagnetic domains can be separated by a new type of domain wall, i.e., a stripe of antiferromagnetic insulating phase\cite{37,38,39}. Based on a standard double exchange model, it was predicted that an insulating domain wall can be energetically lower than a conventional Bloch domain wall, provided that the magnetic anisotropy is sufficiently strong \cite{38}. Unlike the conventional domain wall, the insulating domain wall may still survive even when the spins of the ferromagnetic domains are aligned. Here for the present LPCMO nanowires with extreme aspect ratio, significant magnetic anisotropy is revealed from the magnetic and transport measurements, and can thus stabilize the insulating domain walls between FMM domains under magnetic field up to 14 T, leading to the observed tunneling transport. As illustrated in the previous sections, this picture was also verified by the MFM characterization and magnetic measurement. In contrast, the conventional Bloch domain walls disappear at relative small field, which can be inferred from the close of hysteresis loops at about 0.5 T shown in Figure 1b. We stress here again that the robustness of the insulating domain walls is not caused by the pinning effect from defects, as discussed earlier. Instead, this robustness should be attributed to the strong magnetic anisotropy of the quasi-1D nanowires, in consistent with the theoretical prediction.\cite{38}

Previous studies have revealed that the quantum nature of electrons modifies the classic percolation picture for three-dimensional and two-dimensional electron systems in that the quantum interference plays an important role \cite{40}. Analogously, the quantum behavior of electrons may also alter the classic 1D percolation picture, in which percolative conduction can be only reached when the system is completely in the metallic phase. Here we demonstrate that this classic percolation picture is modified essentially, and a novel quantum percolation state is realized by quantum tunneling across the intrinsic insulating barriers along the quasi-1D transport path. Such quantum percolation state is robust in the low temperature range even under magnetic field up to 14 T, and the revised phase diagram for the LPCMO nanowires is depicted in Figure 3.

In summary, by using single-crystalline LPCMO/MgO core-shell nanowires as a model system, we demonstrate that the quasi-1D confinement on the electronically phase-separated manganites substantially enhances the sensitivity of transport to even probe the magnetic fluctuations, \emph{e}.\emph{g}., the magnetic nano-droplets in the insulating matrix, which is the precursor to the FMM phase. More interestingly, the quasi-1D confinement can even modify the phase competition to stabilize thin insulating domains at low temperatures, which serve as tunneling barriers to form intrinsic tunneling junctions. Such tunneling effect survives even under magnetic field up to 14 T, and essentially modifies the classic 1D percolation picture to stabilize a novel quantum percolation state. A new phase diagram for this model manganite system under quasi-1D confinement is thus established for the first time, which differs substantially from that for the bulk or thin films. Our findings inspire new insight into understanding and manipulation of the EPS and corresponding magneto-transport properties in electronically soft matters via dimensionality control, and thus hold great promises towards electronic device applications.


\section{Extra information}

\subsection{Author Contributions}

$^{\perp}$These authors contributed equally to this work.

\begin{acknowledgement}
This work was supported by National Key Basic Research Program (2014CB921102), National Natural Science Foundation of China (11434009, 11374279, 11461161009 and U1632160), Strategic Priority Research Program (B) of the Chinese Academy of Sciences (XDB01020000), Fundamental Research Funds for the Central Universities (WK2030020027), and National Key Research and Development Plan (2016YFA0401003).
\end{acknowledgement}

\begin{suppinfo}
Supporting Information Available: [Experimental section, magnetic anisotropy of the LPCMO
nanowires, series two-resistor model simulation, magneto-transport and magnetic properties of
the LPCMO thin films, anisotropic tunneling magnetoresistance effects in the LPCMO nanowire and
supplementary figures].
\end{suppinfo}

\section{TOC Graphic}
\begin{figure}
\includegraphics[width=9cm]{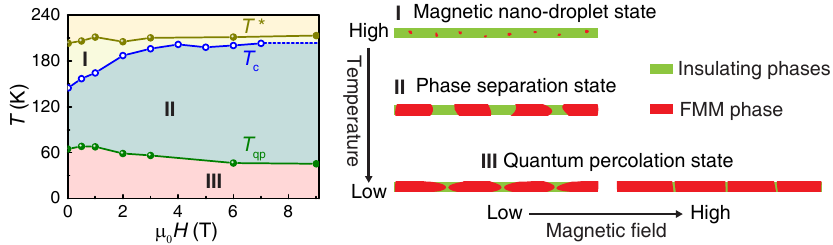}

\end{figure}

\section{Figures}

\begin{figure}
\includegraphics[width=9 cm]{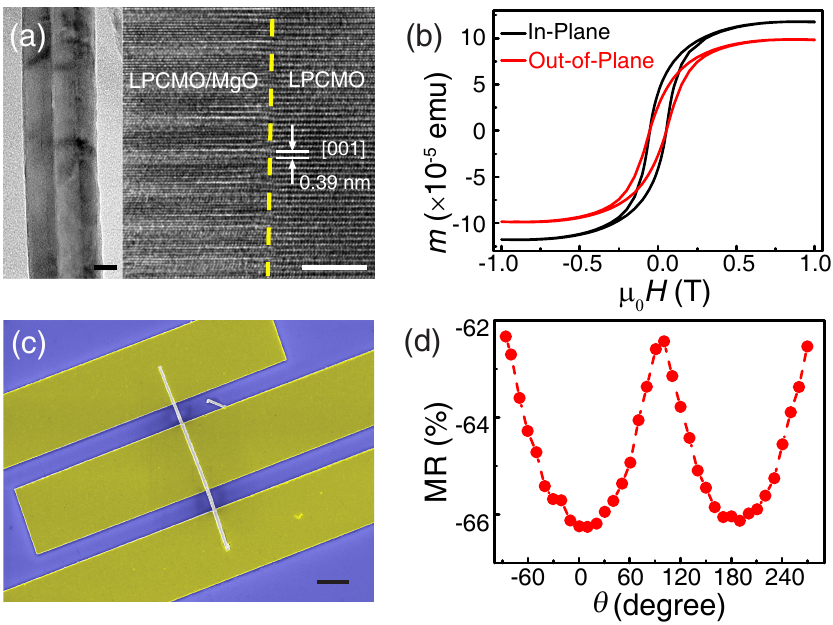}
\caption{\label{fig:figure1} (a) TEM images of a single nanowire. The scale bars in the left and right panels are 20 nm and 5 nm, respectively. (b) Magnetic field dependent magnetic moment at 10 K. The measurements were conducted after zero-field cooling from room temperature. (c) SEM image of a LPCMO nanowire device for transport measurements. The scale bar is 2 $\mu$m. (d) Angle-dependent MR of a LPCMO nanowire measured at 65 K and 0.5 T, where the angle is between the magnetic field and the nanowire long axis.}
\end{figure}

\begin{figure}
\includegraphics[width=\columnwidth]{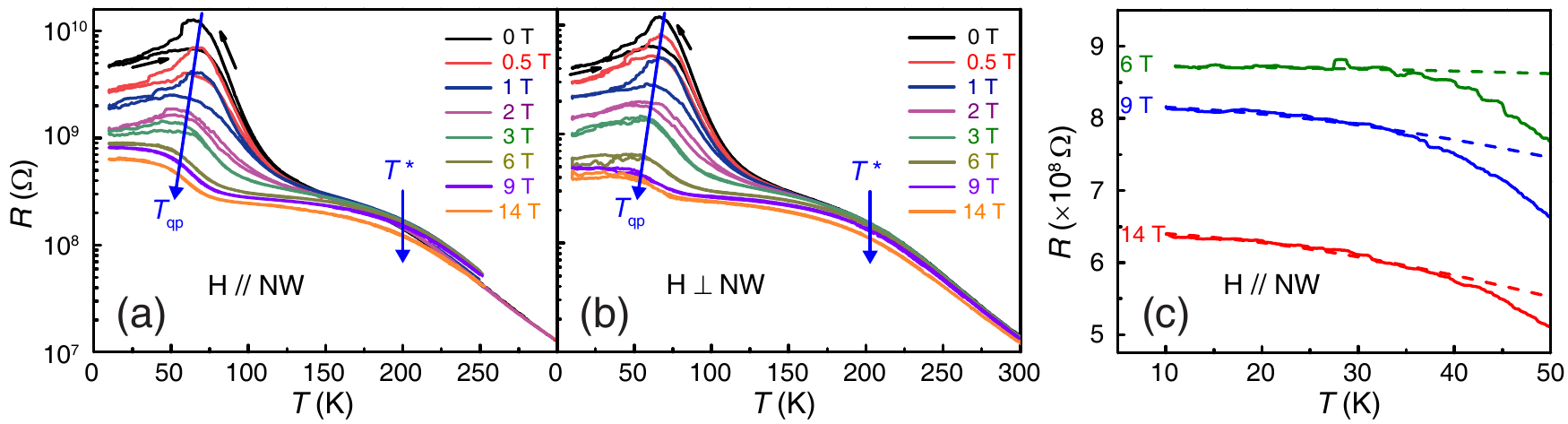}
\caption{\label{fig:figure2} (a,b) \emph{R}-\emph{T} curves at various magnetic fields with the field parallel and perpendicular to the LPCMO nanowire, respectively. (c) Fittings of the \emph{R}-\emph{T} curves at different high magnetic fields.}
\end{figure}

\begin{figure}
\includegraphics[width=15cm]{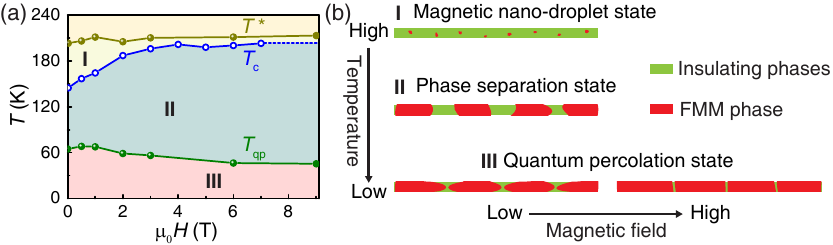}
\caption{\label{fig:figure3} (a) Phase diagram extracted from the magneto-transport measurements (solid circles) and magnetic measurements (open circles). \emph{T}* repesents the onset temperature of magnetic nano-droplets, \emph{T}$_{\rm{c}}$ the Curie temperature, and \emph{T}$_{\rm{qp}}$ the onset temperature of quantum tunneling. Region I refers to the magnetic nano-droplet state, region II refers to the electronic phase separation state, and region III refers to the quantum percolation state. (b) Schematic of the phase evolutions for different states.}
\end{figure}

\begin{figure}
\includegraphics[width=15cm]{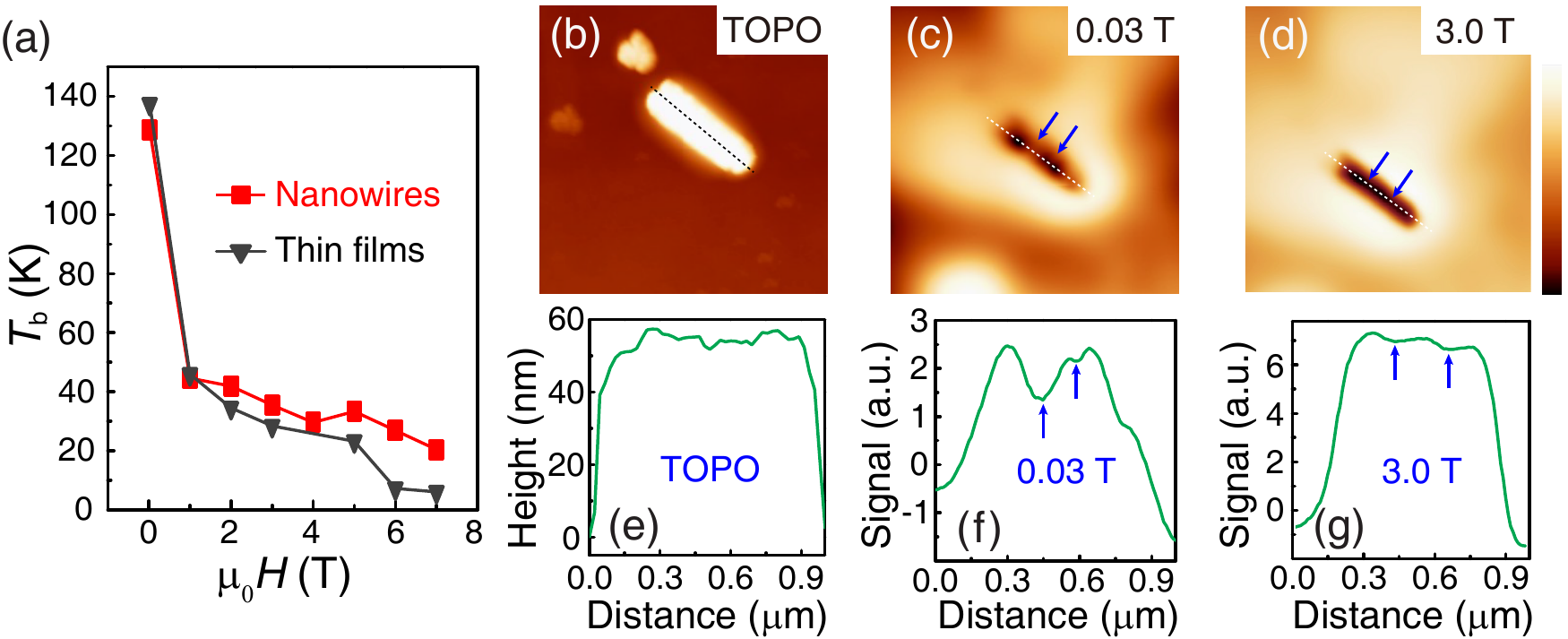}
\caption{\label{fig:figure4} (a) \emph{T}$_{\rm{b}}$ versus magnetic field for the LPCMO nanowires and thin films. (b) Topology image and (c,d) MFM images of a LPCMO nanowire taken at 50 K after zero-field cooling from room temperature. The color scales for (c) and (d) are 4.545 and 10.209 Hz, respectively. Scanned areas in (b-d) are 2.5 $\mu$m $\times$ 2.5 $\mu$m. (e-g) Line profiles along the lines in (b), (c) and (d), respectively.}
\end{figure}

\end{document}